\documentclass[12pt]{iopart}

\usepackage{graphicx}
\usepackage{hyperref}
\usepackage{epstopdf}
\usepackage{amsthm}

\usepackage{array}
\usepackage{bm,bbm}
\usepackage{amssymb}
\usepackage{wrapfig}
\usepackage{caption}
\usepackage{subfigure}
\renewcommand{\thesubfigure}{\thefigure-\arabic{subfigure}}
\makeatletter
\renewcommand{\@thesubfigure}{(\thesubfigure)\space}
\renewcommand{\p@subfigure}{}
      \makeatother
\usepackage{iopams}
\begin{document}

\title{Detector-decoy high-dimensional quantum key distribution}
\author{Hai-ze Bao$^{1,2}$,Wan-su Bao$^{1,2,*}$,Yang Wang$^{1,2}$,Rui-ke Chen$^{1,2}$, Chun Zhou$^{1,2}$, Mu-sheng Jiang$^{1,2}$ and Hong-wei Li $^{1,2}$}
\address{$^1$Zhengzhou Information Science and Technology Institute, Zhengzhou 450001, China}
\address{$^2$Synergetic Innovation Center of Quantum Information and Quantum Physics, University of Science and Technology of China, Hefei, Anhui 230026, China}
\ead{2010thzz@sina.com}

\vspace{10pt}
\begin{indented}
\item[]June 2016
\end{indented}

\begin{abstract}
  The decoy-state high-dimensional quantum key distribution provides a practical secure way to share more private information with high photon-information efficiency. In this paper, based on detector-decoy method, we propose a detector-decoy high-dimensional quantum key distribution protocol. Employing threshold detectors and a variable attenuator, we can estimate single-photon fraction of postselected events and Eve¡¯s Holevo information under the Gaussian collective attack with much simpler operations in practical implementation. By numerical evaluation, we show that without varying source intensity and optimizing decoy-state intensity, our protocol could perform much better than one-decoy-state protocol and as well as the two-decoy-state protocol. Specially, when the detector efficiency is lower, the advantage of the detector-decoy method becomes more prominent.
\end{abstract}

\vspace{2pc}

%\submitto{\NJP}
%\maketitle
\section{Introduction}
Quantum key distribution (QKD) allows two authorized parties, called Alice and Bob, to share private and secure information at a long distance \cite{1,2}. Since BB84 protocol \cite{3} was proposed, lots of work on enhancing the security of QKD has been done, such as measurement-device-independent QKD \cite{Samuel,4,Pirandola,5,6} and round-robin differential phase-shift QKD \cite{7}. Compared with two-level QKD protocols, high-dimensional quantum key distribution (HD-QKD) \cite{8} enables two parties to generate a secret key at a higher rate. By encoding information in a high-dimensional photonic degrees of freedom, such as position-momentum \cite{8}, time-energy \cite{9,10,11,12,13} and orbital angular momentum \cite{14,15,16,17}, HD-QKD can share more information securely per detected single-photon between two parties so as to improve the secret-key capacity under realistic technical constraints. Moreover, HD-QKD protocols could tolerate more noise than qubit QKD protocols \cite{18}.

Recently, based on time-energy entanglement and dispersive optics, a so-called dispersive optic HD-QKD \cite{19} was proposed. It has been proven that the protocol is secure against Gaussian collective attacks \cite{20}. However, like most two-level general QKD protocols, the security of HD-QKD would be influenced by some practical imperfection. Under these imperfect conditions, the imperfect single-photon source always emits multipair events, which causes the HD-QKD vulnerable to the photon number splitting (PNS) attack \cite{21,22,23} over lossy channels.

To avoid the PNS attack, decoy-state method is designed \cite{24} and developed \cite{25,26,27,28,29,30}. The central idea of the decoy-state method is to estimate the channel transmission properties by choosing source intensity settings independently and randomly. Recently, Zhenshen Zhang \emph{et al.} \cite{20} extended decoy-state analysis to HD-QKD protocols with infinite number of decoy states. Darius Bunandar \emph{et al.} \cite{31} proposed a finite decoy-state HD-QKD protocols and its finite-key analysis was presented in Ref\cite{32}.

In this paper, by modifying Alice¡¯s detection setups simply, we introduce the detector-decoy method into the HD-QKD protocol. Compared with the original decoy-state HD-QKD protocol, without varying the source intensity and optimizing decoy-state intensity, the lower bound of single-photon fraction of postselected events and the upper bound of the leaked information can still be obtained in the detector-decoy HD-QKD. The basic idea of the method is to measure the incoming light pulses with a set of detectors with different efficiencies. We use a variable attenuator to change the transmittance of channel in Alice¡¯s side so that the detector efficiency can be changed. Borrowing realistic experimental parameters, we show by numerical evaluations that the detector-decoy HD-QKD could perform much better than the one-decoy-state HD-QKD. Besides, it performs as well as the two-decoy-state HD-QKD protocol, which can perform as well as the HD-QKD protocol with an infinite number of decoy states. Although our method is considered in the DO-QKD, the same arguments are also applicable to other HD-QKD based on time-energy entanglement.

The paper is organized as follows. In Sec. II, we describe the detector-decoy HD-QKD protocol. In Sec. III, we propose a security analysis for HD-QKD. Then, in Sec. IV we show the result of a numerical evaluation with realistic experimental constraints. Finally, section V. concludes the paper with a summary.
\begin{figure}
  \centering
  \includegraphics[width=0.8\textwidth]{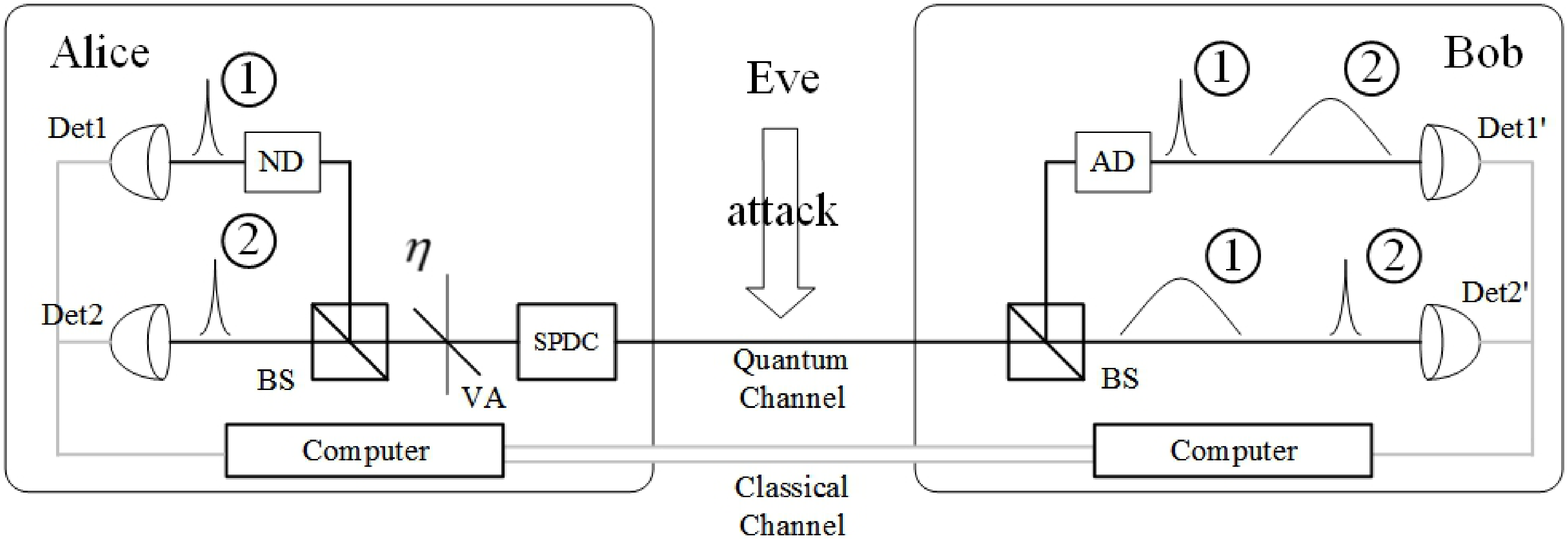}
  \caption{Schematic of the detector-decoy HD-QKD setup. The source is located in Alice¡¯s side. Alice keeps one photon of the pair and sends the other to Bob. They both choose to measure in either the arrival-time basis (case 1) or the frequency basis (case 2) independently and randomly. Their results are only correlated or anti-correlated when they choose the same basis. VA is a variable attenuator. ND is normal dispersion and AD is anomalous dispersion.}
  \label{fig.1}
\end{figure}

\section{Protocol description}
In the original decoy-state protocol \cite{31}, there is an assumption that Eve cannot intrude into Alice and Bob¡¯s experimental setups. This means that Eve just can attack the quantum channel. The source is located in Alice¡¯s side, so we could consider the photon kept by Alice cannot be affected by Eve. The assumption is also needed in this paper.

The detector-decoy method is proposed firstly by Moroder \emph{et al.} \cite{33}, which have showed the advantages and feasibility in QKD. Based on detector-decoy method, we propose a detector-decoy HD-QKD scheme by modifying the Alice¡¯s setup slightly as shown in Fig.~\ref{fig.1}. Specially, our scheme only need one intensity of the source.

A time-energy entangled state could be produced by Alice weakly pumping the SPDC source. We define $\sigma _{cor}$ as the correlation time between two photons, which is determined by the phase matching bandwidth of the SPDC source. Correspondingly, we define $\sigma _{coh}$ as the coherence time of the pump field, which is typically far larger than $\sigma _{cor}$. Thus, the number of alphabet characters per photon pulse, $d={\sigma _{coh}}/{\sigma _{cor}}$ (the Schmidt number) is large \cite{34,35}.

The protocol is described as follow:
\begin{enumerate}
  \item \emph{State preparation:} Alice chooses a transmittance of the VA randomly and independently and then generates a biphoton state by SPDC. She keeps one of the biphoton and sends the other to Bob via a quantum channel.
  \item \emph{State measurement:} Alice selects a basis between time basis and frequency basis randomly and then measures her photon. After receiving the photon from Alice, Bob also performs measurement as Alice and records the outcome.
  \item \emph{Classical information post-processing:} After all signals are transmitted, Alice publishes her transmittance choice and both of them publish their basis choices over an authenticated public channel. They establish their distilled key only from correlated events acquired in the same basis. Alice and Bob would announce some their measurement results to determine postselection probabilities and excess-noise multipliers. Then Eve¡¯s influence could be detected and they can determine their information advantage over Eve. If the advantage is much greater than zero, they then apply error correction and privacy amplification [36] on their data. As a result, some amount of secret key can be established.
\end{enumerate}

\section{Security analysis}
In the decoy-state HD-QKD protocol, the postselection probability can be written as \cite{31}
\begin{center}
\begin{equation}
P_{\mu}=\sum_{n=0}^{\infty}{Pr_n\gamma_n}.
\end{equation}
\end{center}
Here, $\mu$ is the intensity of Alice¡¯s SPDC source and $Pr_n$ is the probability of generating $n-$photon pair. $\gamma_n$ is the conditional probability of measuring at least one detection given $n-$photon pairs are emitted. In our protocol, without Eve¡¯s effect, $\gamma_n$ can be written explicitly as
%{\setlength\abovedisplayskip{0pt}
%\setlength\belowdisplayskip{0pt}
\begin{equation}
\gamma _n=\alpha _{n}^{\left( \eta \right)}\beta _n=\left[ 1-\left( 1-\eta \eta _{Alice} \right) ^n\left( 1-p_d \right) \right] \left[ 1-\left( 1-\eta _{Bob}\eta _T \right) ^n\left( 1-p_d \right) \right]
\end{equation}
Here, $\alpha _{n}^{\left( \eta \right)}=\left[ 1-\left( 1-\eta \eta _{Alice} \right) ^n\left( 1-p_d \right) \right]$ is the conditional probability of Alice registering at least one detection when the transmittance of the variable attenuator is $\eta$ and $\beta _n=\left[ 1-\left( 1-\eta _{Bob}\eta _T \right) ^n\left( 1-p_d \right) \right]$ is the conditional probability of Bob registering at least one detection in a single measurement frame. $\eta _{Alice}$ and $\eta _{Bob}$ are Alice and Bob¡¯s detector efficiencies respectively. $\eta_T$ is is the transmittance of the quantum channel and $p_d$ is the dark count rate in a single measurement frame. Eve, in principle, has the ability to affect the $\gamma _n$ values by Eve¡¯s effect only on $\beta _n$ while $\alpha _{n}^{\left( \eta \right)}$ is cannot be affected due to the assumption.

The bound on the secure-key capacity is \cite{20,25,31,37}
\begin{equation}
\Delta I\geq \beta I\left( A;B \right) -\left( 1-F_{\mu}^{\left( \eta \right)} \right) I_R-F_{\mu}^{\left( \eta \right)}\chi _{\zeta _t,\zeta _{\omega}}^{UB}\left( A;E \right).
\end{equation}
Here,$F_{\mu}^{\left( \eta \right)}=\mu e^{-\lambda}\alpha _{1}^{\left( \eta \right)}\beta _1/{P_{\mu}}$ is the fraction of postselected events that are due to single photon emissions, $\zeta _t$ and $\zeta _{\omega}$ are the value of time excess noise and frequency excess noise respectively, which are caused by multiphoton emissions and dark counts. $\beta$ is reconciliation efficiency and $I\left( A;B \right)$ is the mutual information between Alice and Bob. $I_R$ is the shared information between Alice and Bob. $\chi _{\zeta _t,\zeta _{\omega}}^{UB}\left( A;E \right)$ is an upper bound on Eve¡¯s Holevo information under Gaussian collective attacks due to the excess noise.

In our scheme, for simplicity, we use photon states under the transmittance $\eta_1$ to generate keys. We consider that the statistics of entangled photon-pair produced by SPDC source are approximately Poissonian \cite{38}. So in a single measurement frame, when the mean photon-pair number is $\mu$
\begin{equation}
Pr_n=\frac{\mu ^n}{n!}e^{-\mu}.
\end{equation}

To estimate Eve's effect on $\beta_n$, we request Alice to randomly change the transmittance of the variable attenuator $\eta$ between $\eta_1$ and $\eta_2$ $\left( 0\le \eta _2<\eta _1\le 1 \right) $.

\subsection{Lower bound on $F_{\mu}^{\left( \eta _1 \right)}$}
We define $P_{\mu}^{\left( \eta \right)}$ as the postselection probabilities when the transmittance is $\eta$. The postselection probabilities under the two different transmittance $eta_1$ and $\eta_2$ $\left( 0\le \eta _2<\eta _1\le 1 \right) $ could be written by
\begin{equation}\label{Pr1}
P_{\mu}^{\left( \eta _1 \right)}=\sum_{n=0}^{\infty}{Pr_n\alpha _{n}^{\left( \eta _1 \right)}\beta _n}=\sum_{n=0}^{\infty}{\frac{\mu ^n}{n!}e^{-\mu}\left( 1-\left( 1-\eta _1\eta _{Alice} \right) ^n\left( 1-p_d \right) \right) \beta _n}.
\end{equation}
\begin{equation}\label{Pr2}
P_{\mu}^{\left( \eta _2 \right)}=\sum_{n=0}^{\infty}{Pr_n\alpha _{n}^{\left( \eta _2 \right)}\beta _n}=\sum_{n=0}^{\infty}{\frac{\mu ^n}{n!}e^{-\mu}\left( 1-\left( 1-\eta _2\eta _{Alice} \right) ^n\left( 1-p_d \right) \right) \beta _n}.
\end{equation}
Specially, when $\eta=1$, $P_{\mu}^{\left( \eta \right)}$ is equivalent to $P_{\mu}$ in essence.

Then we will use the Eq.~\ref{Pr1} and Eq.~\ref{Pr2} to deduce a lower bound of $\beta_1$.
\begin{center}
\begin{eqnarray}
&\quad\quad\quad\frac{e^{\mu}P_{\mu}^{\left( \eta _1 \right)}}{1-\left( 1-\eta _1\eta _{Alice} \right) ^2\left( 1-p_d \right)}-\frac{e^{\mu}P_{\mu}^{\left( \eta _2 \right)}}{1-\left( 1-\eta _2\eta _{Alice} \right) ^2\left( 1-p_d \right)}\nonumber\\
&=\beta _0\left[ \frac{p_d}{1-\left( 1-\eta _1\eta _{Alice} \right) ^2\left( 1-p_d \right)}-\frac{p_d}{1-\left( 1-\eta _2\eta _{Alice} \right) ^2\left( 1-p_d \right)} \right]\nonumber\\
&+\beta _1\left[ \frac{1-\left( 1-\eta _1\eta _{Alice} \right) \left( 1-p_d \right)}{1-\left( 1-\eta _1\eta _{Alice} \right) ^2\left( 1-p_d \right)}\mu -\frac{1-\left( 1-\eta _2\eta _{Alice} \right) \left( 1-p_d \right)}{1-\left( 1-\eta _2\eta _{Alice} \right) ^2\left( 1-p_d \right)}\mu \right]\\
&+\sum_{n=3}^{\infty}{\beta _n\left[ \frac{1-\left( 1-\eta _1\eta _{Alice} \right) ^n\left( 1-p_d \right)}{1-\left( 1-\eta _1\eta _{Alice} \right) ^2\left( 1-p_d \right)}\mu -\frac{1-\left( 1-\eta _2\eta _{Alice} \right) ^n\left( 1-p_d \right)}{1-\left( 1-\eta _2\eta _{Alice} \right) ^2\left( 1-p_d \right)}\mu \right]}.\nonumber
\end{eqnarray}
\end{center}
The inequality
\begin{equation}
\frac{1-\left( 1-\eta _1\eta _{Alice} \right) ^n\left( 1-p_d \right)}{1-\left( 1-\eta _2\eta _{Alice} \right) ^n\left( 1-p_d \right)}
\le \frac{1-\left( 1-\eta _1\eta _{Alice} \right) ^2\left( 1-p_d \right)}{1-\left( 1-\eta _2\eta _{Alice} \right) ^2\left( 1-p_d \right)}
\end{equation}
for $n\geq 2$ is satisfied given $0\le \eta _2<\eta _1<1$. That means the equation
\begin{equation}
\label{betaineq}
\fl \qquad\sum_{n=3}^{\infty}{\beta _n\left[ \frac{1-\left( 1-\eta _1\eta _{Alice} \right) ^n\left( 1-p_d \right)}{1-\left( 1-\eta _1\eta _{Alice} \right) ^2\left( 1-p_d \right)}\mu ^n-\frac{1-\left( 1-\eta _2\eta _{Alice} \right) ^n\left( 1-p_d \right)}{1-\left( 1-\eta _2\eta _{Alice} \right) ^2\left( 1-p_d \right)}\mu ^n \right]}\le 0.
\end{equation}

Then Eq.~\ref{betaineq} leads to the following inequality:
\begin{equation}
\fl\beta _1\geq \beta _{1}^{LB}=\frac{\frac{e^{\mu}P_{\mu}^{\left( \eta _1 \right)}}{1-\left( 1-\eta _1\eta _{Alice} \right) ^2\left( 1-p_d \right)}-\frac{e^{\mu}P_{\mu}^{\left( \eta _2 \right)}}{1-\left( 1-\eta _2\eta _{Alice} \right) ^2\left( 1-p_d \right)}+\beta _0\left[ \frac{p_d}{1-\left( 1-\eta _1\eta _{Alice} \right) ^2\left( 1-p_d \right)}-\frac{p_d}{1-\left( 1-\eta _2\eta _{Alice} \right) ^2\left( 1-p_d \right)} \right]}{\frac{1-\left( 1-\eta _1\eta _{Alice} \right) \left( 1-p_d \right)}{1-\left( 1-\eta _1\eta _{Alice} \right) ^2\left( 1-p_d \right)}-\frac{1-\left( 1-\eta _2\eta _{Alice} \right) \left( 1-p_d \right)}{1-\left( 1-\eta _2\eta _{Alice} \right) ^2\left( 1-p_d \right)}}.
\end{equation}

Now we need to lower bound $\beta_0$. As it¡¯s shown in Ref\cite{20,31}, under the assumption that Eve cannot intrude into both experimental setups, the probability $\beta_0$ cannot be lower than the dark count rate no matter what Eve does:
\begin{equation}
\beta _0\geq \beta _{0}^{LB}=p_d.
\end{equation}
Therefore, we can give rise to the fraction of postselected events that are due to single photon emissions as:
\begin{eqnarray}\label{Fu}
\fl&F_{\mu}^{\left( \eta _1 \right)}=\alpha _{1}^{\left( \eta _1 \right)}\beta _1\frac{\mu e^{-\mu}}{P_{\mu}^{\left( \eta _1 \right)}}\geq \left( 1-\left( 1-\eta _1\eta _{Alice} \right) \left( 1-p_d \right) \right) \\
\fl&\times\frac{\frac{P_{\mu}^{\left( \eta _1 \right)}}{1-\left( 1-\eta _1\eta _{Alice} \right) ^2\left( 1-p_d \right)}-\frac{P_{\mu}^{\left( \eta _2 \right)}}{1-\left( 1-\eta _2\eta _{Alice} \right) ^2\left( 1-p_d \right)}+e^{-\mu}\beta _{0}^{LB}\left[ \frac{p_d}{1-\left( 1-\eta _1\eta _{Alice} \right) ^2\left( 1-p_d \right)}-\frac{p_d}{1-\left( 1-\eta _2\eta _{Alice} \right) ^2\left( 1-p_d \right)} \right]}{P_{\mu}^{\left( \eta _1 \right)}\left[ \frac{1-\left( 1-\eta _1\eta _{Alice} \right) \left( 1-p_d \right)}{1-\left( 1-\eta _1\eta _{Alice} \right) ^2\left( 1-p_d \right)}-\frac{1-\left( 1-\eta _2\eta _{Alice} \right) \left( 1-p_d \right)}{1-\left( 1-\eta _2\eta _{Alice} \right) ^2\left( 1-p_d \right)} \right]}\nonumber
\end{eqnarray}

\subsection{Lower bound on $\zeta _t$ and $\zeta _\omega$}
When Eve attacks Alice¡¯s transmission, the decrease in measurement correlations of Alice and Bob is caused, which is parameterized by the excess-noise factors $\zeta _t$ and $\zeta _\omega$.

We consider the relation in Ref.\cite{31}. $\Omega _x$ (for $x=t$ and $\omega$) is the averaged excess-noise multiplier, which can be measured by two authenticated parties and used to estimate $\zeta _t$ and $\zeta _\omega$ by following equations:
\begin{equation}\label{multiplier}
\Omega _x=F_{\mu}^{\left( \eta \right)}\left( 1+\zeta _x \right) +\Delta \Omega _x\left( 1-F_{\mu}^{\left( \eta \right)} \right)
\end{equation}
where $x=t$ and $\omega$.

The Eq.~\ref{multiplier} can be diveded into two groups by transmittance $\eta_1$ and $\eta_2$, which can be expressed as:
\begin{eqnarray}
&\Omega _{x,\eta _1}=F_{\mu}^{\left( \eta _1 \right)}\left( 1+\zeta _x \right) +\Delta \Omega _x\left( 1-F_{\mu}^{\left( \eta _1 \right)} \right),\\
&\Omega _{x,\eta _2}=F_{\mu}^{\left( \eta _2 \right)}\left( 1+\zeta _x \right) +\Delta \Omega _x\left( 1-F_{\mu}^{\left( \eta _2 \right)} \right).
\end{eqnarray}
We multiply these two equations by $P_{\mu}^{\left( \eta _1 \right)}e^\mu$ and $P_{\mu}^{\left( \eta _2 \right)}e^\mu$ respectively, and then we have
\begin{eqnarray}
\label{m1}\Omega _{x,\eta _1}P_{\mu}^{\left( \eta _1 \right)}e^\mu=\mu \alpha _{1}^{\left( \eta _1 \right)}\beta _1\left( 1+\zeta _x \right)+\Delta \Omega _x\left( P_{\mu}^{\left( \eta _1 \right)}e^\mu-\mu \alpha _{1}^{\left( \eta _1 \right)}\beta _1 \right),\\
\label{m2}\Omega _{x,\eta _2}P_{\mu}^{\left( \eta _2 \right)}e^\mu=\mu \alpha _{1}^{\left( \eta _2 \right)}\beta _1\left( 1+\zeta _x \right)+\Delta \Omega _x\left( P_{\mu}^{\left( \eta _2 \right)}e^\mu-\mu \alpha _{1}^{\left( \eta _2 \right)}\beta _1 \right).
\end{eqnarray}
Combing Eq.~\ref{m1} and Eq.~\ref{m2}, we get
\begin{eqnarray}
\fl&\qquad\Omega _{x,\eta _1}P_{\mu}^{\left( \eta _1 \right)}e^\mu-\Omega _{x,\eta _2}P_{\mu}^{\left( \eta _2 \right)}e^\mu\nonumber\\
\fl&=\left( \alpha _{1}^{\left( \eta _1 \right)}-\alpha _{1}^{\left( \eta _2 \right)} \right) \mu \beta _1\left( 1+\zeta _x \right) +\Delta \Omega _x\left[ P_{\mu}^{\left( \eta _1 \right)}e^u-P_{\mu}^{\left( \eta _2 \right)}e^u-\left( \alpha _{1}^{\left( \eta _1 \right)}-\alpha _{1}^{\left( \eta _2 \right)} \right) \mu \beta _1 \right]\\
\fl&=\left( \eta _1-\eta _2 \right) \eta _{Alice}\left( 1-p_d \right) \mu \beta _1\left( 1+\zeta _x \right) +\Delta \Omega _x\left[ P_{\mu}^{\left( \eta _1 \right)}e^u-P_{\mu}^{\left( \eta _2 \right)}e^u-\left( \eta _1-\eta _2 \right) \eta _{Alice}\left( 1-p_d \right) \mu \beta _1 \right]\nonumber\\
\fl&\geq \left( \eta _1-\eta _2 \right) \eta _{Alice}\left( 1-p_d \right) \mu \beta _1\left( 1+\zeta _x \right),\nonumber
\end{eqnarray}
where the inequality comes from
\begin{eqnarray}
 P_{\mu}^{\left( \eta _1 \right)}-P_{\mu}^{\left( \eta _2 \right)}=&\sum_{n=0}^{\infty}{\left\{ \left[ \left( 1-\eta _2\eta _{Alice} \right)^n-\left( 1-\eta _1\eta _{Alice} \right) ^n \right]\left( 1-p_d \right) \right\}}\beta _1\nonumber\\
&\geq \left( \eta _1-\eta _2 \right) \eta _{Alice}\left( 1-p_d \right) \mu \beta _1.
\end{eqnarray}
Thus,
\begin{equation}
\fl \quad\qquad \left( 1+\zeta _x \right) \le \frac{\left( \Omega _{x,\eta _1}P_{\mu}^{\left( \eta _1 \right)}-\Omega _{x,\eta _2}P_{\mu}^{\left( \eta _2 \right)} \right) e^{\mu}}{\left( \eta _1-\eta _2 \right) \eta _{Alice}\left( 1-p_d \right) \mu \beta _1}\le \frac{\left( \Omega _{x,\eta _1}P_{\mu}^{\left( \eta _1 \right)}-\Omega _{x,\eta _2}P_{\mu}^{\left( \eta _2 \right)} \right) e^{\mu}}{\left( \eta _1-\eta _2 \right) \eta _{Alice}\left( 1-p_d \right) \mu \beta _{1}^{LB}}
\end{equation}
for $x=t$ and $\omega$.

The upper bounds on $\zeta _t$ and $\zeta _\omega$ could be obtained from:
\begin{eqnarray}
\Omega _{x,\eta}&=F_{\mu}^{\left( \eta \right)}\left( 1+\zeta _x \right) +\Delta \Omega _x\left( 1-F_{\mu}^{\left( \eta \right)} \right)\geq F_{\mu}^{\left( \eta \right)}\left( 1+\zeta _x \right)\nonumber\\
&=\frac{\left[ 1-\left( 1-\eta \eta _{Alice} \right) \left( 1-p_d \right) \right] P_{\mu}^{\left( \eta _1 \right)}}{\left[ 1-\left( 1-\eta _1\eta _{Alice} \right) \left( 1-p_d \right) \right] P_{\mu}^{\left( \eta \right)}}F_{\mu}^{\left( \eta _1 \right)}\left( 1+\zeta _x \right)\\
&\geq \frac{\left[ 1-\left( 1-\eta \eta _{Alice} \right) \left( 1-p_d \right) \right] P_{\mu}^{\left( \eta _1 \right)}}{\left[ 1-\left( 1-\eta _1\eta _{Alice} \right) \left( 1-p_d \right) \right] P_{\mu}^{\left( \eta \right)}}F_{\mu}^{LB}\left( 1+\zeta _x \right)\nonumber
\end{eqnarray}
for $\eta=\eta_1$ or $\eta_2$. This inequality implies that
\begin{equation}
\left( 1+\zeta _x \right) \le \frac{\left[ 1-\left( 1-\eta _1\eta _{Alice} \right) \left( 1-p_d \right) \right] P_{\mu}^{\left( \eta \right)}\Omega _{x,\eta}}{\left[ 1-\left( 1-\eta \eta _{Alice} \right) \left( 1-p_d \right) \right] P_{\mu}^{\left( \eta _1 \right)}F_{\mu}^{LB}}
\end{equation}

Therefore, we can give the upper bounds on $\zeta _t$ and $\zeta _\omega$ as follows:
\begin{eqnarray}\label{noise}
\zeta _x\le \zeta _{x}^{UB}&=\min \left\{ \frac{\left( \Omega _{x,\eta _1}P_{\mu}^{\left( \eta _1 \right)}-\Omega _{x,\eta _2}P_{\mu}^{\left( \eta _2 \right)} \right) e^{\mu}}{\left( \eta _1-\eta _2 \right) \eta _{Alice}\left( 1-p_d \right) \mu \beta _{1}^{LB}},\right.\nonumber\\
& {{\mathop{\min}}\atop{\scriptstyle\eta \in \left\{ \eta _1,\eta _2 \right\}}}\left.\left\{ \frac{\left[ 1-\left( 1-\eta _1\eta _{Alice} \right) \left( 1-p_d \right) \right] P_{\mu}^{\left( \eta \right)}\Omega _{x,\eta}}{\left[ 1-\left( 1-\eta \eta _{Alice} \right) \left( 1-p_d \right) \right] P_{\mu}^{\left( \eta _1 \right)}F_{\mu}^{LB}} \right\} \right\} -1.
\end{eqnarray}

Now with Eq.~\ref{Fu} and Eq.~\ref{noise}, we obtain a lower bound on $F_\mu$ and upper bound on $\zeta _t$ and $\zeta _\omega$.

\section{Numerical evaluation}
We borrow some realistic experimental parameters in the implementation of the HD-QKD \cite{39}: propagation loss $\alpha=0.2$ dB$/$km, detector timing jitter $\sigma _{J}=20$ ps, dark count rate $R_{dc}=1000$ s$^{-1}$, reconciliation efficiency $\beta=0.9$, $n_R=\log _2d$. Therefore, we could know the channel transmittance by the function $\eta _T=10^{{-\alpha L}/{10}}$. The detector efficiencies of Alice and Bob¡¯s setups are assumed to be equal. Here we consider a specific threshold detector, the superconducting nanowire single-photon detector, with the high detection efficiencies $\eta_{Alice}=\eta_{Bob}=0.93$ \cite{40}.

We take coherent time $\sigma _{cor}=30$ ps for all $d$ values and correspondingly $\sigma _{coh}=d\sigma _{cor}$, which is easily controlled . We choose a single measurement frame duration $T_f=2\sqrt{\ln 2}\sigma _{coh}$ \cite{20}. For simplicity, we consider two excess-noise factors are equal, $\zeta_t=\zeta_\omega=\zeta$. The change in correlation time due to Eve¡¯s interaction can be assumed to be $\Delta \sigma =\left( \sqrt{1+\zeta}-1 \right) \sigma _{cor}=10$ ps \cite{41}, which doesn¡¯t lose generality.

In our detector-decoy method, we vary the transmittance of the variable attenuator between $\eta_1$ and $\eta_2$, specifically $\eta_1=1$ and $\eta_2=0.5$. Here, we assume that events in which both Alice and Bob click when $\eta_1=1$ are used to encode information.

\begin{figure}
  \centering
  \subfigure[]{\label{fig:2-1}\includegraphics[width=3in]{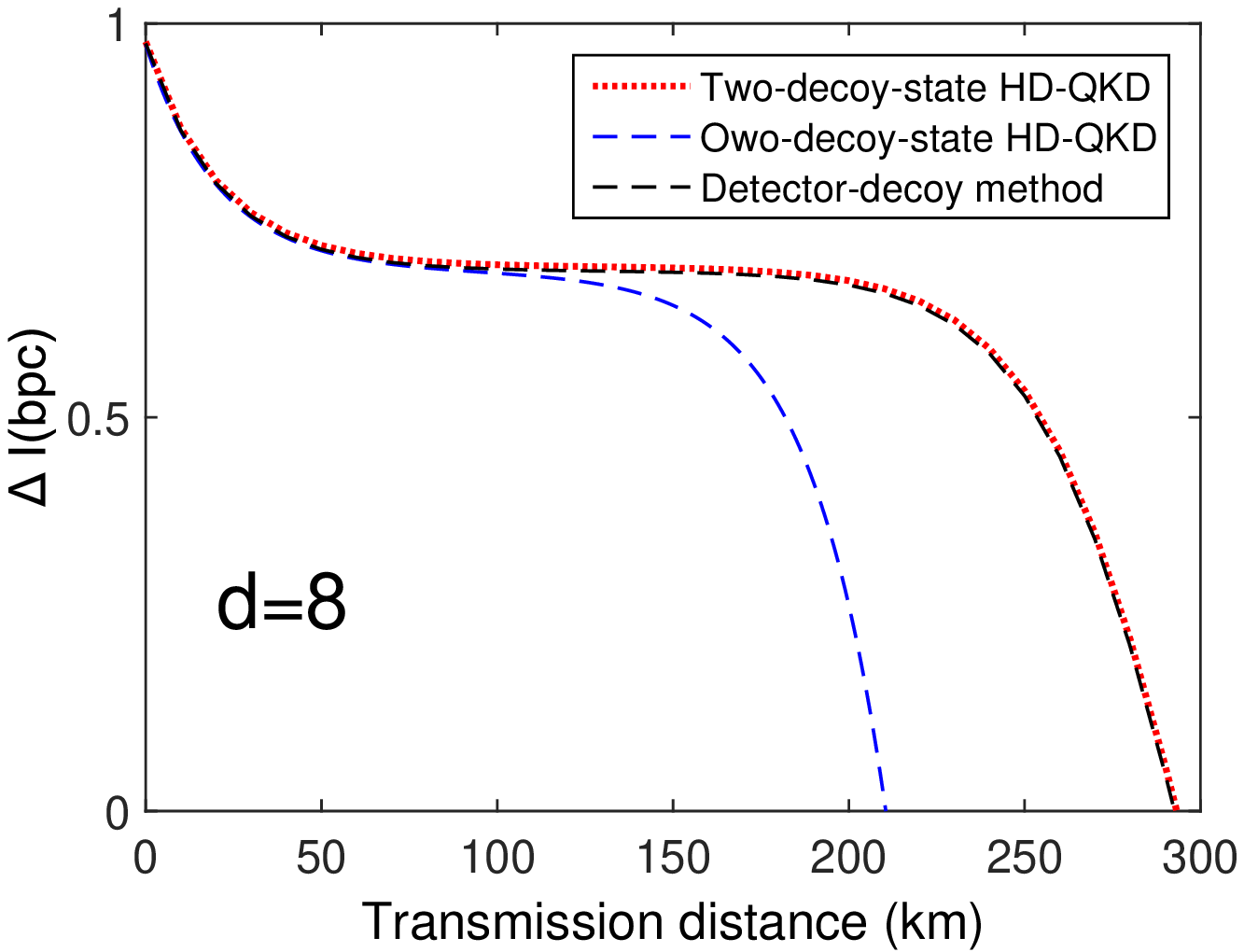}}
  \subfigure[]{\label{fig:2-2}\includegraphics[width=3in]{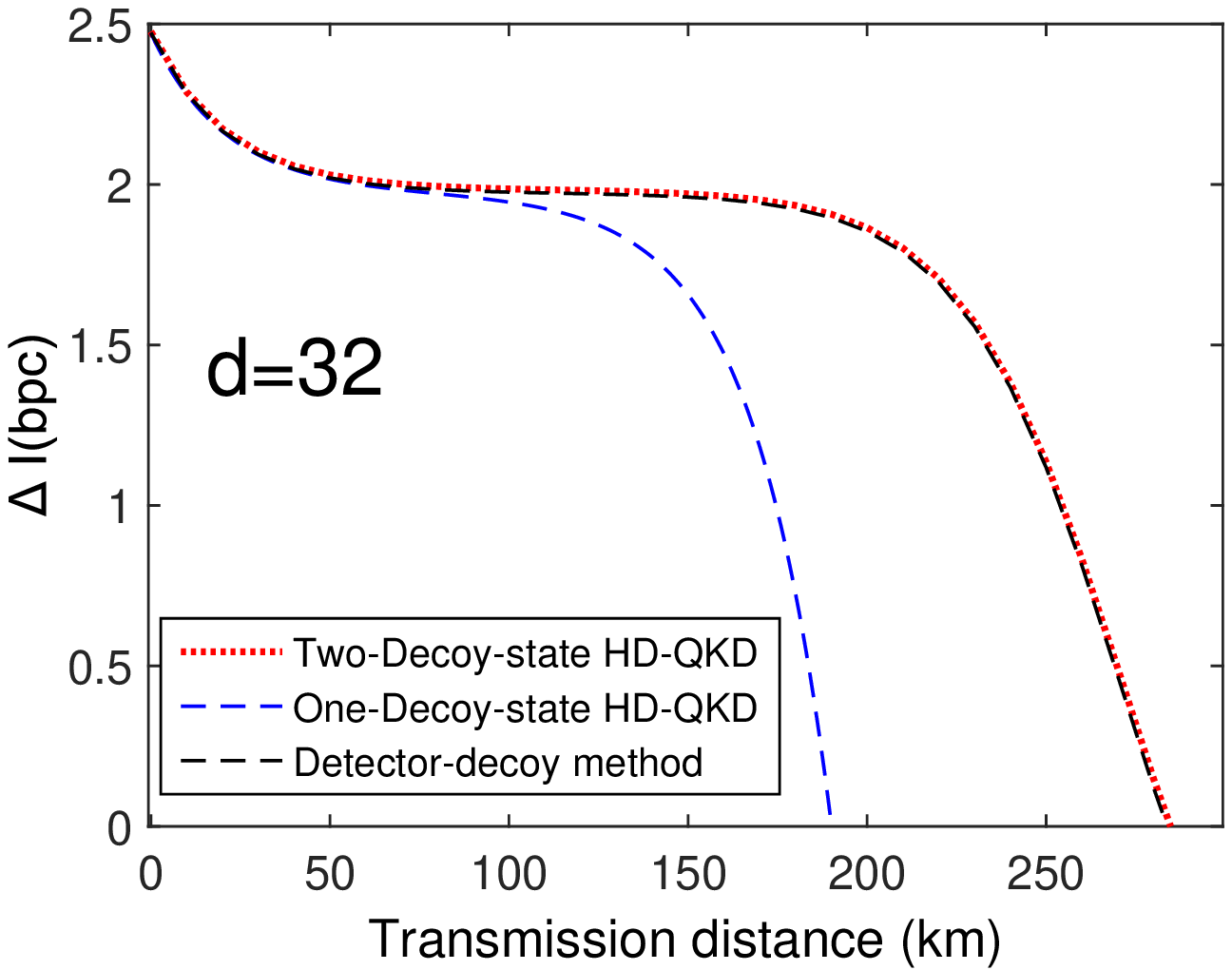}}
  \caption{(Color online) Lower bounds on the secure-key capacity in bits per coincidence for (2-1) $d=8$  and (2-2) $d=32$  as a function of transmission distance at 10 km increments. The results show that the detector-decoy method (black dash lines) could perform much better than one-decoy-state HD-QKD (blue dash lines) and as well as the two-decoy-state HD-QKD (red dotted lines). }
  \label{fig2}
\end{figure}

Here, we give a comparison between the two-decoy-state HD-QKD protocol with average photon number $\mu=0.1$ when the dimension $d=8$ and $d=32$, which is shown in Fig.~\ref{fig2} As one can see, the detector-decoy method can perform much better than one-decoy-state HD-QKD and as well as the two-decoy-state HD-QKD protocol, which means that the detector-decoy HD-QKD could work as the decoy-state HD-QKD with infinite number of decoy states without the source intensity modulation or monitoring photon numbers \cite{32}.

\begin{figure}
  \centering
  \subfigure[]{
    \label{fig:3-1}
    \includegraphics[width=3in]{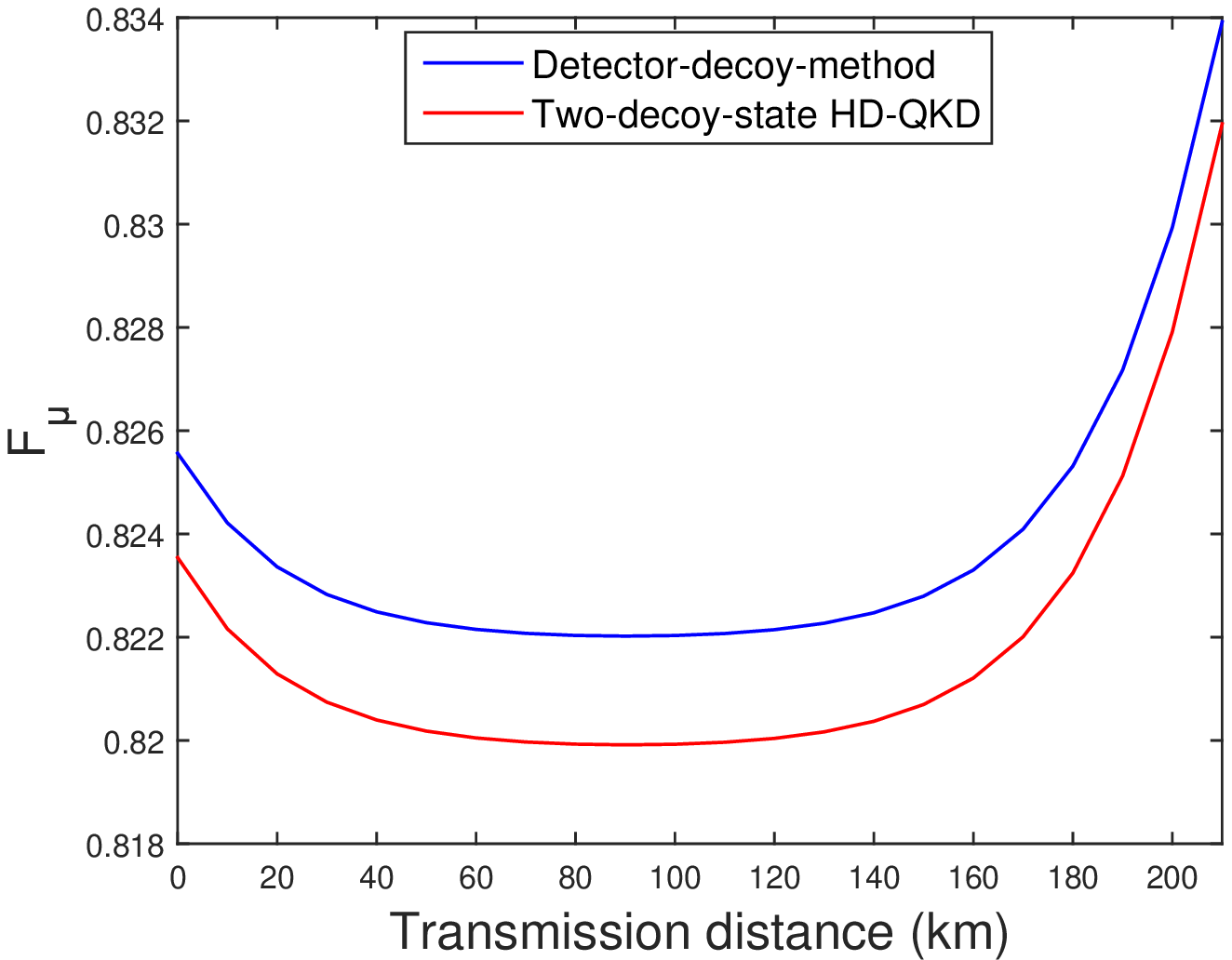}}
  %\hspace{0in}
  \subfigure[]{
    \label{fig:3-2}
    \includegraphics[width=3in]{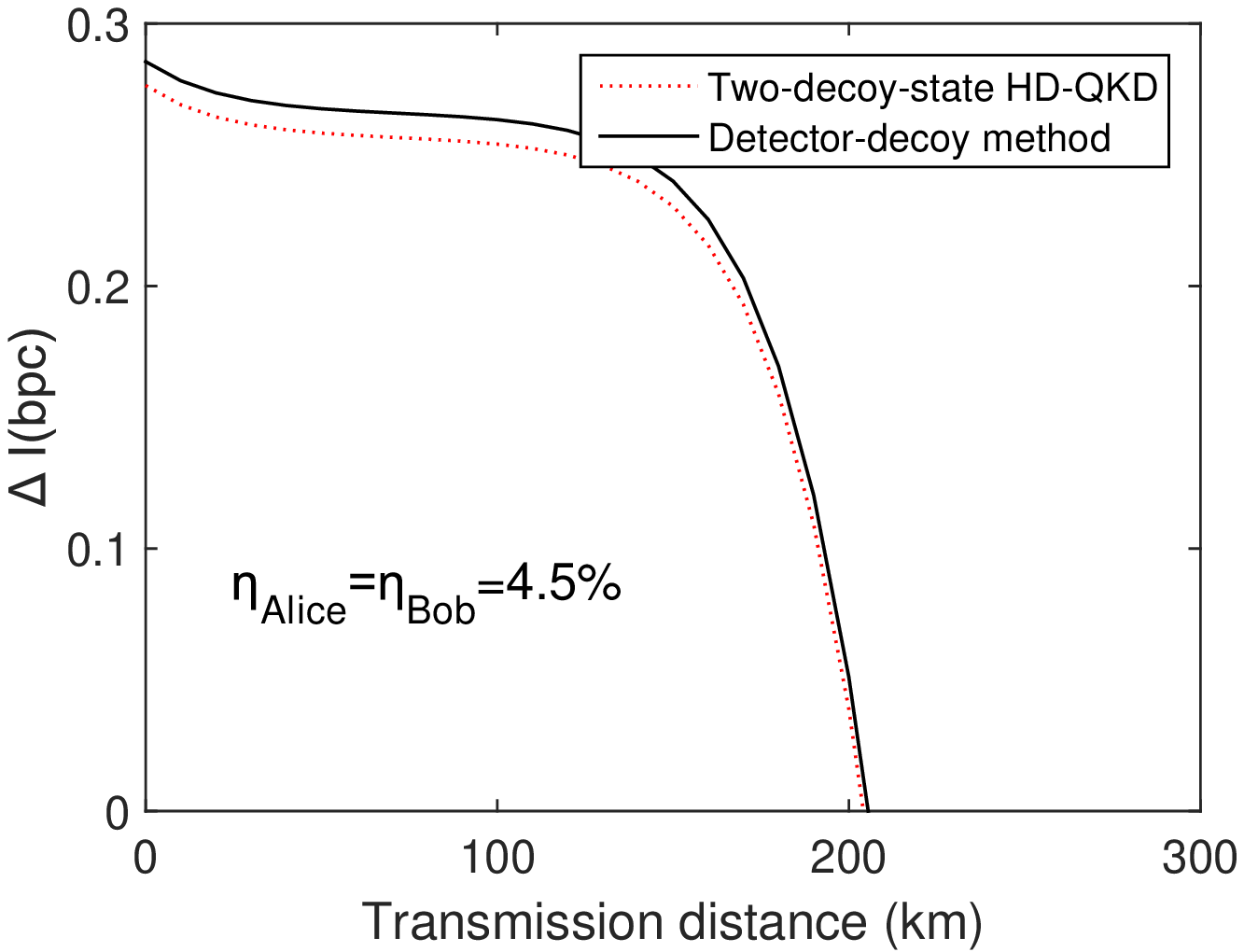}}
  \caption{(Color online) (3-1) The $F_\mu$ values obtained by detector-decoy method (blue solid lines) and two-decoy-state HD-QKD (red solid lines). (3-2) The secret key capacity per coincidence for original decoy-state HD-QKD (red dotted lines) and detector-decoy HD-QKD (black solid lines) as a function of the transmission distance when  $d=8$, $\mu=0.10$ and $\eta_{Alice}=\eta_{Bob}=0.045$.}
  \label{fig3}
\end{figure}

In general, the efficiency of conventional detectors used in QKD experiment is always far lower than the efficiency of the superconducting nanowire single-photon detector \cite{42}.So, we also give a comparison between schemes in the condition that is limited with low detector efficiencies, especially $\eta_{Alice}=\eta_{Bob}=0.045$ shown in Fig.~\ref{fig3}. Fig.~\ref{fig:3-1} plots the $F_\mu$ values that are obtained by detector-decoy method (blue solid lines) and two-decoy-state HD-QKD (red solid lines) respectively. Fig.~\ref{fig:3-2} shows the comparison between two protocols. Both of them are plotted under the condition with $d=8$ and $\mu=0.10$.As the figure shown, when the detector efficiency is low, the advantage of detector-decoy HD-QKD become prominent. That means the detector-decoy HD-QKD could tolerate lower detector efficiency.

\section{Conclusion}
The decoy-state method is applied to resist PNS attacks in HD-QKD systems. The original decoy-state HD-QKD has proved that the two-decoy-state protocol can perform as well as the protocol with infinite decoy states. In this paper, we proposed a detector-decoy HD-QKD protocol that could offer certain practical advantages over the original decoy-state HD-QKD protocol. Our scheme, using a variable attenuator, could achieve secure information per coincidence as the two-decoy-state HD-QKD with little or no compromise. In the two-decoy-state protocol, the intensity of weaker decoy states is optimized for any particular transmission distance. However, in our scheme the source intensity is constant and not required to be varied in our scheme. This is a main advantage over the original scheme. Specially, if the detector efficiency is lower, the advantage of our method become more prominent.

In realistic constraints, the detector-decoy HD-QKD presented in this paper provides a more applicable and feasible scheme in the practical implementation of the HD-QKD protocol with slight modification. The source intensity in our scheme is constant and correspondingly we do not need to optimize the intensity of decoy states. When the HD-QKD protocol with decoy states is implemented practically, our scheme can perform as well as the original two-decoy-state HD-QKD protocol but operations are much simpler.

\ack
This work was supported by the National Basic Research Program of China (Grant No. 2013CB338002) and the National Natural Science Foundation of China (Grants No. 11304397 and No. 61505261).

\end{document}